\documentclass{article}
\usepackage[hyphens]{url}
\usepackage[square,sort,comma,numbers]{natbib}
\usepackage{amsmath}
\usepackage{graphicx}
\usepackage{caption}
\usepackage{subcaption}
\usepackage{csquotes}
\usepackage{array}
\usepackage{threeparttable}
\usepackage{authblk}
\usepackage{fancyhdr}

\usepackage{forest}

\tikzset{
Above/.style={
  midway,
  above,
  font=\scriptsize,
  text width=1.5cm,
  align=center,
  },
Below/.style={
  midway,
  below,
  font=\scriptsize,
  text width=1.5cm,
  align=center
  }
}

\title{Do Public Events Affect Sex Trafficking Activity?}
\date{February 15, 2016}

\author[1]{Kyle Miller\thanks{mille856@andrew.cmu.edu}}
\author[1]{Emily Kennedy\thanks{emilyk@alumni.cmu.edu}}
\author[1]{Artur Dubrawski\thanks{awd@cs.cmu.edu}}
\affil[1]{Auton Lab, Carnegie Mellon University}

\fancypagestyle{plain}{
\fancyfoot[C]{\footnotesize Copyright \textcopyright\ 2016 Kyle Miller, Emily Kennedy, Artur Dubrawski}
}

\begin{document}
\maketitle

\begin{abstract}
For several years the pervasive belief that the Super Bowl is the single biggest day for human trafficking in the United States each year has been perpetuated in popular press despite a lack of evidentiary support. The practice of relying on hearsay and popular belief for decision-making may result in misappropriation of resources in anti-trafficking efforts. We propose a data-driven approach to analyzing sex trafficking, especially as it is carried on during--and perhaps in response to--large public events such as the Super Bowl. We examine 33 public events, chosen for attendance numbers comparable to the Super Bowl from a diversity of types, and use the volume of escort advertisements posted online as an accessible and reasonable proxy measure for the actual levels of activity of sex-workers as well as trafficking victims. Our analysis puts the impact of local public events on sex advertisement activity into perspective. We find that many of the events we considered are not correlated with statistically significant impact on sex-worker advertising, though some are. Additionally, we demonstrate how our method can uncover evidence of other events, not included in our initial list, that are correlated with more significant increases in ad activity. 
Reliance on quantitative evidence accessible through data-driven analysis can inform wise resource allocation, guide good policies, and foster the most meaningful impact.  

\end{abstract}

\section{Introduction}

\begin{displayquote}
``The Super Bowl is the greatest show on Earth, but it also has an ugly underbelly.  It's commonly known as the single largest human trafficking incident in the United States.''~\cite{USAToday}
\end{displayquote}
\begin{displayquote}
``...the dirty little secret is that the Super Bowl actually is one of the highest levels of human sex trafficking activity of any event in the country.''~\cite{WashPost}
\end{displayquote}
\begin{displayquote}
``The Super Bowl has become one of the largest venues for sex trafficking in the country.''~\cite{Klobuchar}
\end{displayquote}
Quotes like these have heavily influenced public understanding of the problem of human trafficking in recent years.  Yet, there has been no discernible evidentiary basis for the claim that the Super Bowl is the biggest day for human trafficking in the United States each year.  Support has been largely anecdotal~\cite{HuffPost}. Many opponents of this belief have pointed out that there is a lack of data to establish the hypothetical connection between the Super Bowl and sex trafficking~\cite{ASUstudy}. Relying on hearsay and popular belief for decision-making may result in misappropriation of resources in anti-trafficking efforts. 
Attempting to alleviate this issue, we propose a data-driven approach to analyzing sex trafficking, especially as it is carried on during--and perhaps in response to--large public events such as the Super Bowl.  

We consider a broad sample of annual festivals, sporting events, conventions and and similar events,
and we use the volume of escort advertisements posted online as an accessible and reasonable proxy measure for the actual levels of activity of sex-workers as well as trafficking victims. These levels are difficult to measure directly~\cite{dank2014estimating,kennedy2012predictive}. We examine 33 public events, chosen for attendance numbers comparable to the Super Bowl from a diversity of types, in an attempt to quantify evidence of their potential impact on trafficking. In addition, we conduct a comprehensive analysis of advertisement trends to identify spikes in sex ad activity, and by the assumed implication, spikes in sex trafficking activity.

Our analysis puts the impact of local public events on sex advertisement activity into perspective. We find that many of the events we considered are not correlated with statistically significant impact on sex-worker advertising, while evidence of others that are correlated was uncovered by the analysis. We hope that these types of analyses will give the public, researchers, lawmakers, and law enforcement agencies an improved understanding of the problem of sex trafficking in North America as it relates to large scale events and conventions.

\section{Methods}

We analyze female escort advertisements scraped from popular classifieds websites. This data set represents advertisements collected between October 2011 and February 2016 published for specific United States locations and between July 2013 and February 2016 for locations in Canada. At present, this data set comprises upwards of 32 million advertisements. However, due to complexities involving scraping frequency, network load, etc.\ the scraping procedure does not capture all advertisements at all times. Thus, we make use of a simple bivariate temporal anomaly detection algorithm~\cite{dubrawski2011detection} that includes the notion of a baseline to help control for some of these issues. The approach is to construct a $2\times 2$ contingency table using a pair of sliding windows over time series of total ad counts for a location compared to data for the state in which it is contained. We use a 7-day analysis window immediately preceded by a 91-day reference window. Fisher's exact test then gives the expected ad count for the location in the analysis window and the significance of deviation from expectation. Dates reported correspond to the end-date of the 7-day analysis window.

\begin{figure}
\centering
{\footnotesize
\begin{forest} 
for tree={grow=east,draw=blue, circle,line width=0.2pt,parent anchor=east,child anchor=west,
  edge={draw=blue},edge label={\Huge\color{black}},
  edge path={\noexpand\path[\forestoption{edge}]
      (!u.parent anchor) -- ([xshift=-1.6cm].child anchor) --    
      (.child anchor)\forestoption{edge label};}, l sep=2cm,} 
[,rectangle, s sep=35pt,
  [First,edge label={node[Below]{No related ads posted on any prior day}}
  ]
  [,edge label={node[Below]{Otherwise}}
    [Local,edge label={node[Below]{Related ad posted in same location within 7 days}}
    ]
    [,edge label={node[Below]{Otherwise}}
      [Local,edge label={node[Below]{Most recent related ad posted in same location}}
      ]
      [New-to-town,edge label={node[Below]{Otherwise}}
      ]
    ]
  ]
]
\end{forest}}
\caption{Decision tree illustrating the definition of `new-to-town' ad classification.}
\label{ntt:decision_tree}
\end{figure}
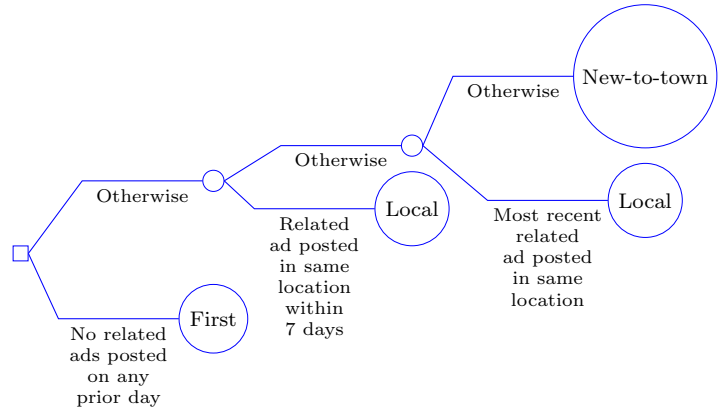

The event-driven sex trafficking narrative purports that events draw traffickers and their victims from disparate locations to the event vicinity. To measure evidence of such behavior it is necessary to identify which advertisements represent individuals recently arrived from elsewhere. We refer to this concept as `new-to-town.' We identify `new-to-town' ads by grouping advertisements into equivalence classes according to shared phone numbers (a distinct identifier present in ~95\% of ads). Then, an ad is identified as representing a `first appearance' group if no related ads were posted on a day prior to the given ad. Otherwise, an ad is identified as representing `local' activity if either its location is among the most recent locations within the related ads or a related ad was posted in the same location within the previous 7 days. If an ad is not identified as representing a `first appearance' or `local' activity, then it is identified as `new-to-town'. Figure~\ref{ntt:decision_tree} shows a decision tree representing this classification scheme. `New-to-town' ads make up 8\% of the data and `first appearance' ads make up 6.5\%, on average. It should be noted that thusly defined`new-to-town' behavior is likely underestimated, and 'first appearance' consequently overestimated, due to the common practice of using "burner phones," i.e.\ periodically changing one's phone number, a strategy used in the sex industry as well as among sex traffickers~\cite{latonero2012technology}. Phone number extraction is accomplished using a regular expression based information extractor written specifically to identify phone numbers in ad text~\cite{dubrawski2015leveraging}.

Using the temporal scan algorithm, we report the most statistically significant exceedance in ad volume in the period spanning 7 days prior to the start to 7 days following the end of each event. Exceedance is defined as the difference between the number of ads expected based on historical trends and number of actual ads observed.  Significance is reported as empirical p-value: percentage of all sliding windows which are more significant. Temporal scan was used to analyze anomalousness of both the total ad volume and the `new-to-town' ad volume respectively, which would help us identify unusual spikes in the data that may be correlated with an event.

Additionally, we applied the temporal scan algorithm using the configurations described above for `new-to-town' ad volume comprehensively to all locations for the calendar year of 2015. We identified the top 3 most significant deviations overall, that is those that represented the most extreme evidence of a large influx of new-to-town sex-workers in our data, and potentially of sex trafficking victims. Interestingly, these top 3 signals were not included in the list of specific events we traced as described above.  The 3 signals were included in the summary of event specific analyses to provide additional perspective.
%

\section{Results}

\begin{figure}
\centering
\includegraphics[width=0.75\linewidth]{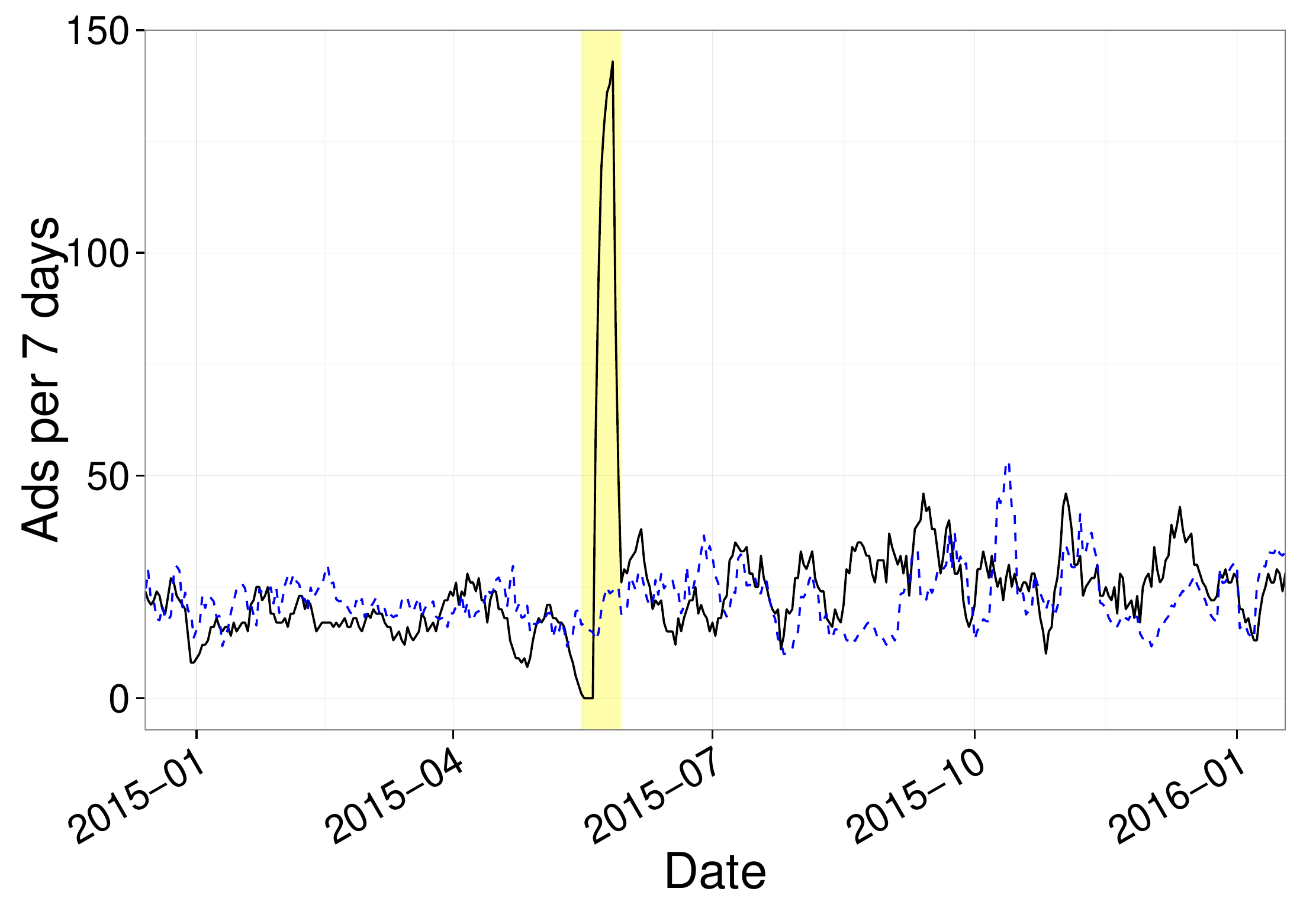}
\caption{Time series of `new-to-town' ad volume for Vancouver, British Columbia, Canada. Black (solid) line represents observed volume,  7-day sum moving at daily resolution. Blue (dashed) line represents expected volume. Highlight represents most significant exceedance $\pm 7$ days. Highlight is centered at 2015-05-23.}
\label{temporalscanplots:1}
\end{figure}

\begin{figure}
\centering
\includegraphics[width=0.75\linewidth]{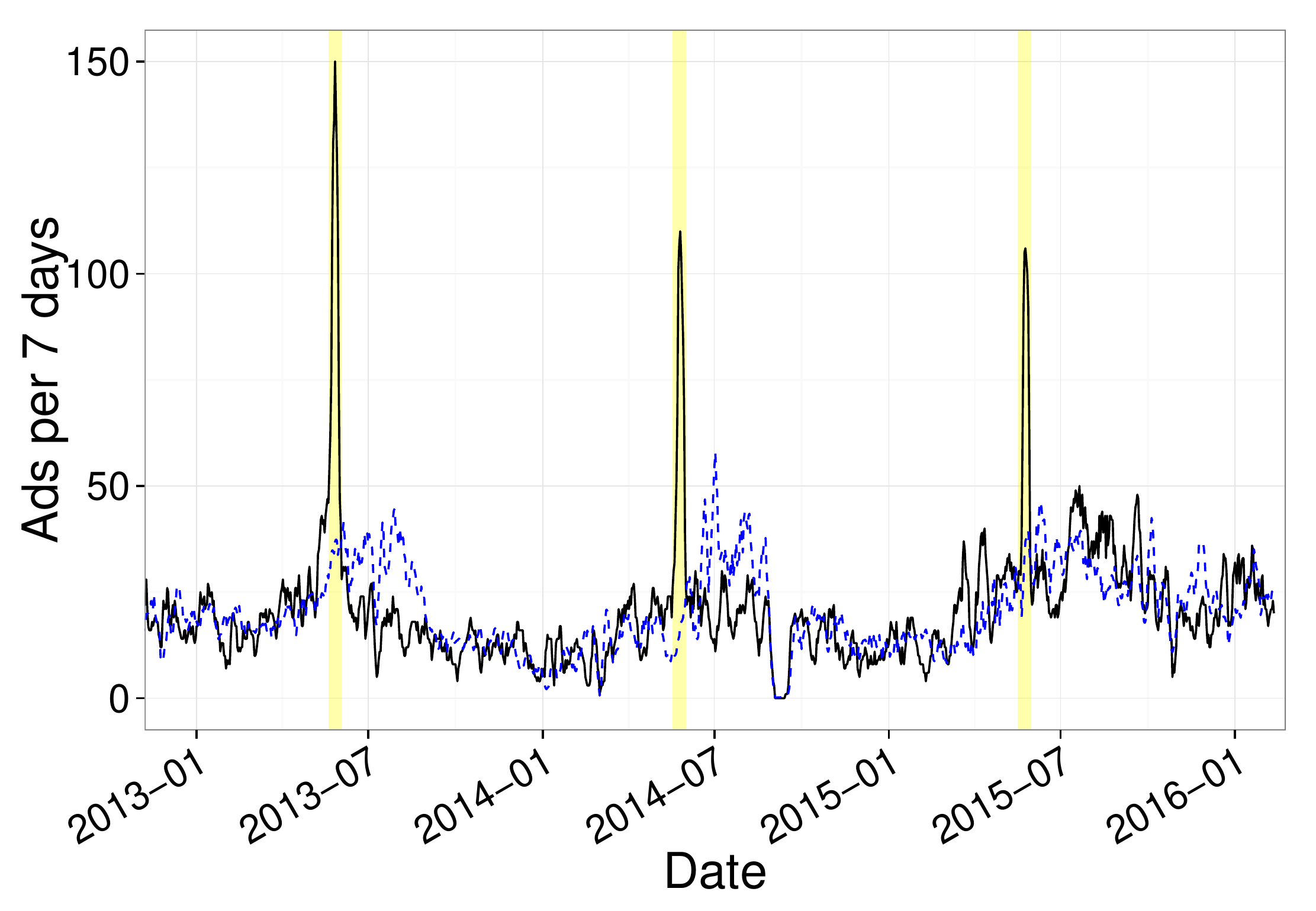}
\caption{Time series of `new-to-town' ad volume for Myrtle Beach, South Carolina, USA. Black (solid) line represents observed volume,  7-day sum moving at daily resolution. Blue (dashed) line represents expected volume. Highlight represents most significant exceedance within each calendar year $\pm 7$ days. Highlights are centered at 2013-05-24, 2014-05-25, and 2015-05-27.}
\label{temporalscanplots:2}
\end{figure}

\begin{figure}
\centering
\includegraphics[width=0.75\linewidth]{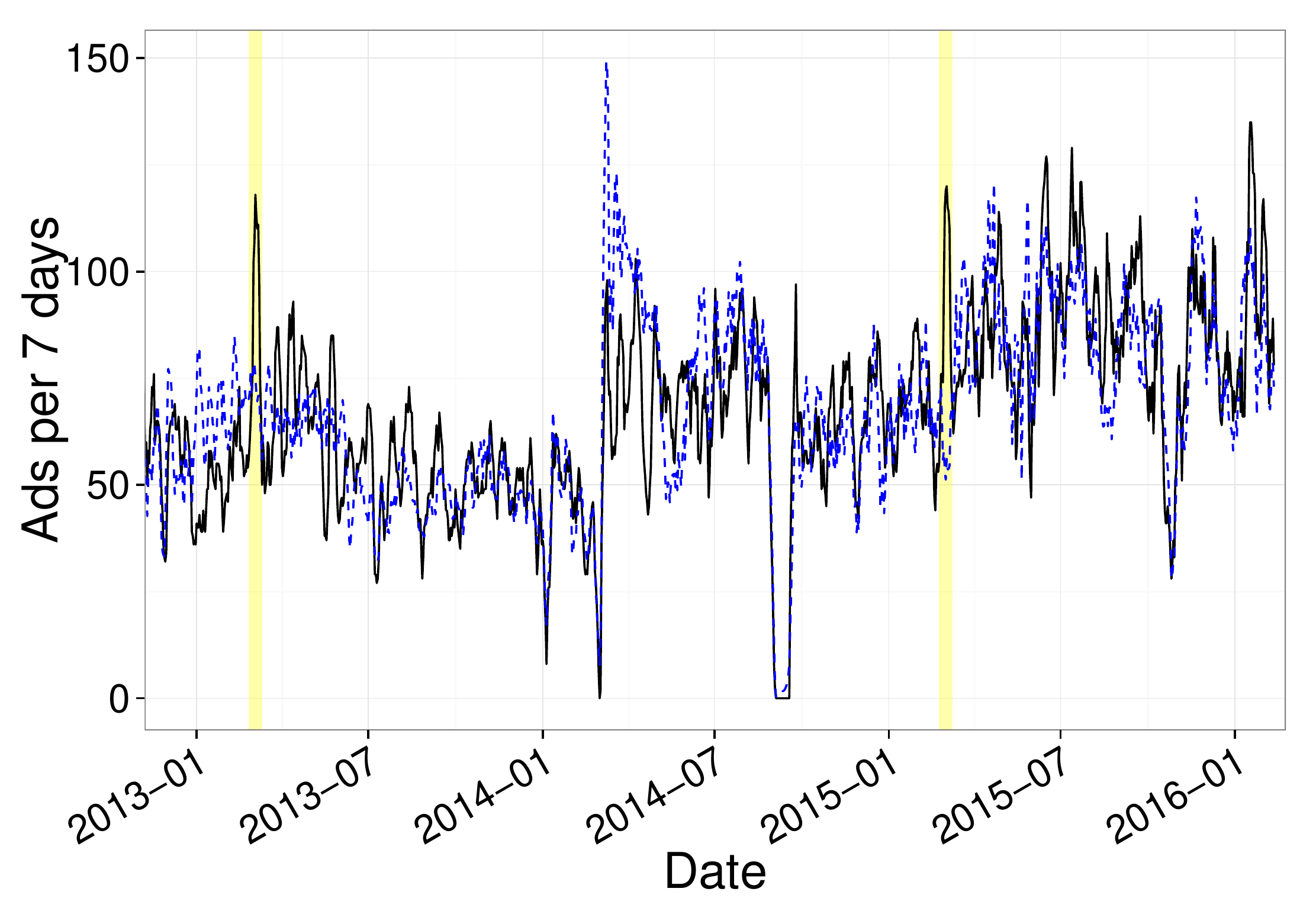}
\caption{Time series of `new-to-town' ad volume for Charlotte, North Carolina, USA. Black (solid) line represents observed volume,  7-day sum moving at daily resolution. Blue (dashed) line represents expected volume. Highlight represents most significant exceedance within each calendar year $\pm 7$ days. Highlights are centered at 2013-03-04 and 2015-03-02.}
\label{temporalscanplots:3}
\end{figure}

The three most significant increases in `new-to-town' ad volume were, in order of significance; Vancouver, British Columbia (2015-05-23), Myrtle Beach, South Carolina (2015-05-27), and Charlotte, North Carolina (2015-03-02). Figures~\ref{temporalscanplots:1}, \ref{temporalscanplots:2}, and \ref{temporalscanplots:3} show time series of `new-to-town' ad volumes in a 7-day moving sum for these locations, respectively. Additionally, we determined the most `new-to-town' increases in prior years for these locations. Myrtle Beach showed significant increases on 2013-05-24 and 2014-05-25. Charlotte showed a significant increase on 2013-03-04.

Beyond the top three identified exceedances in the `new-to-town' group, activity quickly falls to within the typical variance observed over time. This is already evident in Figure~\ref{temporalscanplots:3}. While these other exceedances may still be statistically significant, and in fact meaningful, they do not amass compelling evidence of a large scale influx of sex-workers or trafficking victims.

\begin{figure}
\centering
\includegraphics[width=1.0\linewidth]{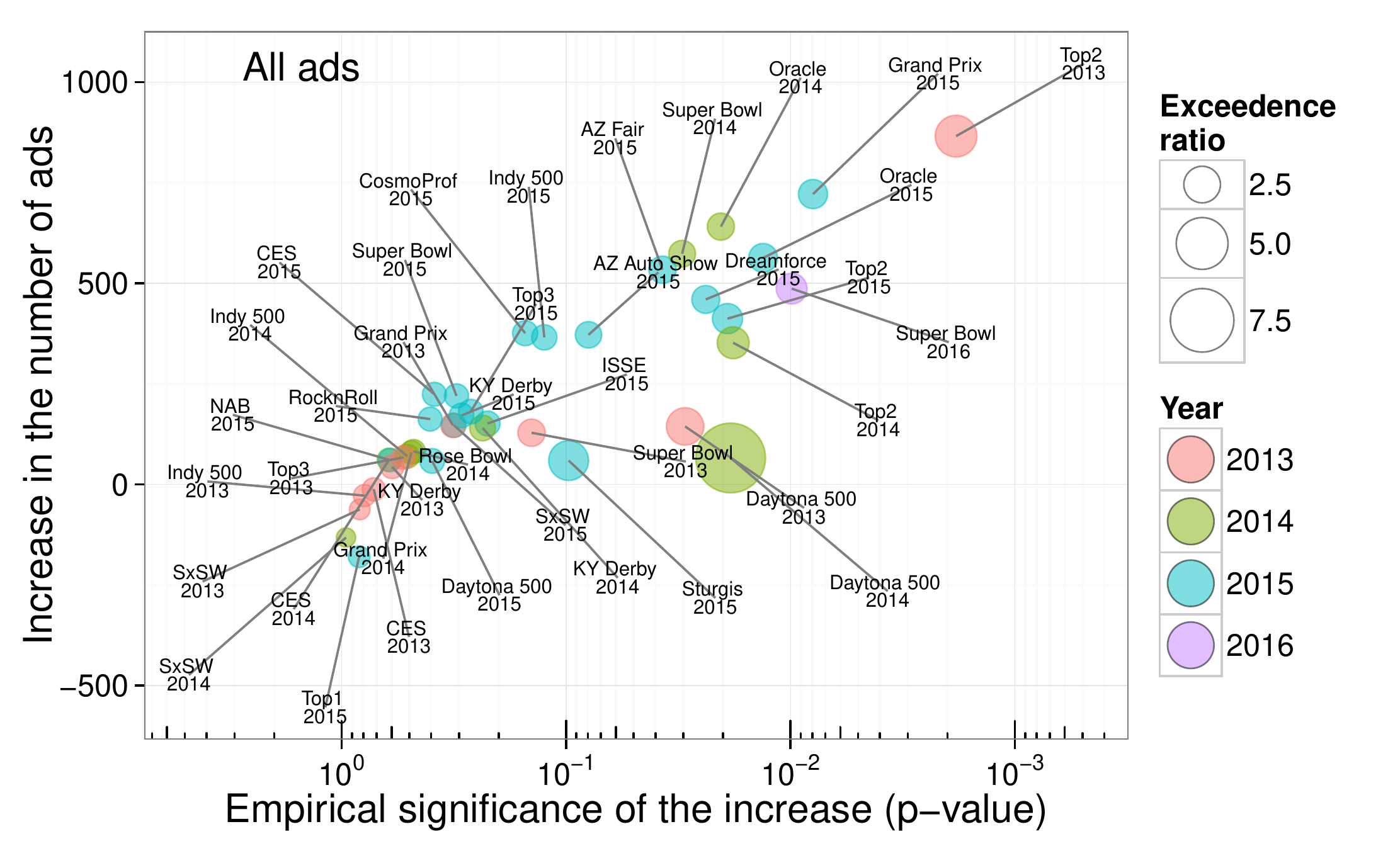}
\caption{Significance of deviation from expected ad volume shown against magnitude of temporary increase in ad volume for all ads posted in the location and time frame of noted events. Point size represents the ratio of observed ad volume to its expectation. Ad counts represent 7-day totals.}
\label{bubbleplots:overall}
\end{figure}

\begin{figure}
\centering
\includegraphics[width=1.0\linewidth]{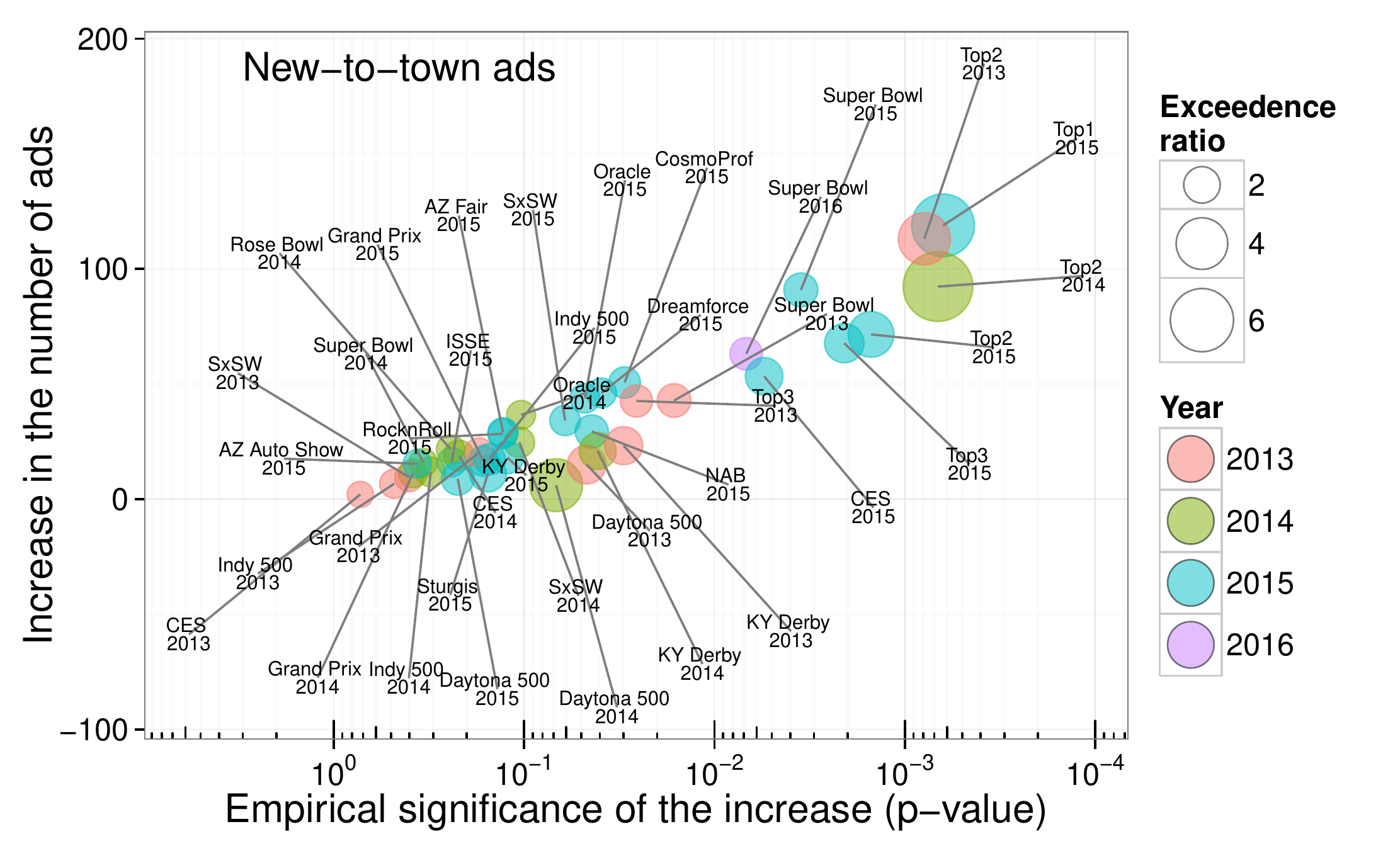}
\caption{Significance of deviation from expected ad volume shown against magnitude of temporary increase in ad volume for `new-to-town' ads posted in the location and time frame of noted events. Point size represents the ratio of observed ad volume to its expectation. Ad counts represent 7-day totals.}
\label{bubbleplots:ntt}
\end{figure}

The Super Bowl events, except for 2014, are on par with the other events that were correlated with the most significant increase in `new-to-town' ads analyzed in this study. The 2015 Consumer Electronics Show (CES) is similar to the Super Bowl events in this regard. The Formula One Grand Prix, Super Bowl 50, and Oracle OpenWorld Convention are correlated with the most significant increases in overall ad volume. Table~\ref{table:allresults} summarizes expected counts and observed counts of published escort ads with the most significant 7-day window around each event, for both `new-to-town' and overall advertising activity. Figure~\ref{bubbleplots:overall} describes the position of the events listed in Table~\ref{table:allresults} relative to statistical significance of deviation from expected total ad volume and the magnitude of the deviation.  Event bubbles that appear higher on the vertical axis in the graph had a larger increase in activity than was expected.  Event bubbles that appear farther to the right on the graph had a more statistically significant increase in activity.  The ratio of the number of ads expected to the number of ads observed is represented by the size of each bubble in the plot: bigger bubbles represent a larger disparity in the number of ads that were expected versus the number that were observed.  
Similarly, Figure~\ref{bubbleplots:ntt} describes the changes in `new-to-town' ad volume correlated with each event. Labels have been abbreviated for presentation, but the assignment of each point ought to be clear from context.

\begin{table}[ht]
\centering
\caption{Expected and observed ad counts for each event for both `new-to-town' ads and overall. P-values given are empirical. Rows are sorted according to the significance of the exceedance in `new-to-town' ads. }
\label{table:allresults}
\resizebox{\linewidth}{!}{
\begin{threeparttable}
\begin{tabular}{>{\raggedright\arraybackslash}p{0.4\linewidth}>{\raggedright\arraybackslash}p{0.3\linewidth}>{\raggedright\arraybackslash}p{0.175\linewidth}cccccc}

 & & & \multicolumn{3}{c}{`New-to-town' ads} & \multicolumn{3}{c}{All ads} \\
Event & Location\tnote{$\dagger$} & Date(s) & expected & observed & p-value & expected & observed & p-value \\ \hline
Top screening result & Vancouver, British Columbia & 2015-05-23 & 24 & 143 & 6.3E-4 & 1409 & 1229 & 8.4E-1 \\
2$^{\rm nd}$ screening result & Myrtle Beach, South Carolina & 2014-05-25 & 15 & 107 & 6.7E-4 & 395 & 747 & 1.8E-2 \\
2$^{\rm nd}$ screening result & Myrtle Beach, South Carolina & 2013-05-24 & 37 & 150 & 7.9E-4 & 386 & 1252 & 1.8E-3 \\
2$^{\rm nd}$ screening result & Myrtle Beach, South Carolina & 2015-05-27 & 33 & 105 & 1.5E-3 & 587 & 999 & 1.9E-2 \\
3$^{\rm rd}$ screening result & Charlotte, North Carolina & 2015-03-02 & 51 & 119 & 2.1E-3 & 1548 & 1729 & 2.7E-1 \\
Super Bowl XLIX & Phoenix, Arizona & 2015-02-01 & 124 & 215 & 3.5E-3 & 2334 & 2554 & 3.1E-1  \\
Consumer Electronics Show (CES) & Las Vegas, Nevada & 2015-01-06 to 2015-01-09 & 48 & 101 & 5.5E-3 & 3636 & 3860 & 3.9E-1 \\
Super Bowl 50 & San Jose, California & 2016-02-07 & 106 & 169 & 6.8E-3 & 624 & 1111 & 9.9E-3 \\
Super Bowl XLVII & New Orleans, Louisiana & 2013-02-03 & 61 & 104 & 1.6E-2 & 325 & 454 & 1.4E-1 \\
3$^{\rm rd}$ screening result & Charlotte, North Carolina & 2013-03-04 & 75 & 118 & 2.6E-2 & 1387 & 1454 & 5.3E-1 \\
CosmoProf & Las Vegas, Nevada & 2015-07-12 to 2015-07-14 & 105 & 156 & 3.0E-2 & 2213 & 2589 & 1.5E-1 \\
Kentucky Derby & Louisville, Kentucky & 2013-05-04 & 20 & 43 & 3.0E-2 & 1103 & 1147 & 6.0E-1 \\
Dreamforce & San Francisco, California & 2015-09-15 to 2015-09-18 & 137 & 183 & 3.9E-2 & 1030 & 1490 & 2.4E-2  \\
Kentucky Derby & Louisville, Kentucky & 2014-05-03 & 20 & 41 & 4.1E-2 & 619 & 759 & 2.4E-1 \\
NAB & Las Vegas, Nevada & 2015-04-11 to 2015-04-16 & 45 & 74 & 4.4E-2 & 2751 & 2812 & 6.1E-1 \\
Daytona 500 & Daytona, Florida & 2013-02-24 & 12 & 27 & 4.7E-2 & 90 & 234 & 3.0E-2 \\
Oracle OpenWorld Convention & San Francisco, California & 2015-10-25 to 2015-10-29 & 135 & 179 & 4.8E-2 & 1029 & 1592 & 1.3E-2 \\
South by Southwest & Austin, Texas & 2015-03-13 to 2015-03-22 & 86 & 120 & 6.1E-2 & 1578 & 1725 & 3.2E-1  \\
Daytona 500 & Daytona, Florida & 2014-02-23 & 2 & 8 & 6.8E-2 & 8 & 74 & 1.8E-2  \\
Oracle OpenWorld Convention & San Francisco,California & 2014-09-28 to 2014-10-02 & 141 & 178 & 1.0E-1 & 1736 & 2377 & 2.0E-2 \\
South by Southwest & Austin, Texas & 2014-03-07 to 2014-03-16 & 61 & 86 & 1.1E-1 & 429 & 297 & 9.6E-1 \\
Kentucky Derby & Louisville, Kentucky & 2015-05-02 & 31 & 49 & 1.2E-1 & 1315 & 1486 & 2.9E-1 \\
Rock n Roll Marathon & Phoenix, Arizona & 2015-01-16 to 2015-01-18 & 86 & 114 & 1.3E-1 & 2399 & 2561 & 4.0E-1 \\
Arizona State Fair & Phoenix, Arizona & 2015-10-16 to 2015-11-08 & 83 & 112 & 1.3E-1 & 1344 & 1878 & 3.7E-2 \\
Sturgis Motorcycle Rally & Rapid City, South Dakota & 2015-08-03 to 2015-08-09 & 11 & 22 & 1.5E-1 & 31 & 90 & 9.7E-2 \\
Indy 500 & Indianapolis, Indiana & 2015-05-24 & 37 & 54 & 1.5E-1 & 1957 & 2323 & 1.3E-1 \\
Formula One United States Grand Prix & Austin, Texas & 2015-10-25 & 40 & 57 & 1.6E-1 & 1212 & 1934 & 7.9E-3 \\
Formula One United States Grand Prix & Austin, Texas & 2013-11-17 & 61 & 81 & 1.7E-1 & 1522 & 1669 & 3.2E-1 \\
Consumer Electronics Show (CES) & Las Vegas, Nevada & 2014-01-07 to 2014-01-10 & 42 & 61 & 2.2E-1 & 2896 & 2957 & 6.2E-1 \\
Daytona 500 & Daytona, Florida & 2015-02-22 & 14 & 23 & 2.2E-1 & 437 & 495 & 4.0E-1  \\
Pro Beauty Conference (ISSE) & Long Beach, California & 2015-01-24 to 2015-01-26 & 58 & 74 & 2.4E-1 & 979 & 1130 & 2.2E-1 \\
Rose Bowl & Los Angeles, California & 2014-01-06 & 103 & 125 & 2.4E-1 & 1470 & 1552 & 4.8E-1 \\
Indy 500 & Indianapolis, Indiana & 2014-05-25 & 35 & 47 & 3.1E-1 & 1111 & 1181 & 5.1E-1 \\
Super Bowl XLVIII & Manhattan, New York & 2014-02-02 & 84 & 100 & 3.4E-1 & 1663 & 2236 & 3.0E-2 \\
Arizona International Auto Show & Phoenix, Arizona & 2015-11-26 to 2015-11-29 & 74 & 89 & 3.6E-1 & 1250 & 1622 & 7.9E-2 \\
Formula One United States Grand Prix & Austin, Texas & 2014-11-02 & 51 & 62 & 3.8E-1 & 1539 & 1618 & 4.9E-1 \\
South by Southwest & Austin, Texas & 2013-03-08 to 2013-03-17 & 38 & 48 & 4.0E-1 & 299 & 237 & 8.3E-1 \\
Indy 500 & Indianapolis, Indiana & 2013-05-26 & 23 & 30 & 4.8E-1 & 289 & 261 & 7.9E-1 \\
Consumer Electronics Show (CES) & Las Vegas, Nevada & 2013-01-08 to 2013-01-11 & 59 & 61 & 7.2E-1 & 2101 & 2089 & 7.2E-1 
\end{tabular}
\begin{tablenotes}
\item[$\dagger$] Location noted refers to regional websites nearest to actual event location.
\end{tablenotes}
\end{threeparttable}}
\end{table}%

\section{Conclusion}

Our analysis has shown that in some cases such as the Super Bowl (in particular, of 2013, 2015, and 2016) or Consumer Electronics Show (2015), one can see a correlation between the occurrence of the event and noticeable and statistically significant evidence of an influx of sex-workers, and, we may imply, sex trafficking activity.  However, in using publicly available data to analyze the broader context of sex trafficking in the United States and Canada, we have identified multiple other venues, events, times, and locations that show a more significant influx of sex advertising than the Super Bowl.  For example, we have identified Memorial Day weekend at Myrtle Beach as systematically demonstrating a large spike in `new-to-town' sex ads, which is more likely to be due to a local event than that seen around the Super Bowl, by an order of magnitude.  The Vancouver instance also shows evidence of a very significant spike suggestive of an influx of sex-workers in response to a local event. 
%
We mention these additional findings not to demonize other events the way that the Super Bowl has been in the past, characterized as the "single largest human trafficking incident in the United States."
Instead, our analysis highlights that human trafficking affects our country across varied locations, communities, and events, and we cannot isolate its impact to only one event per year.

In this work, we focused our attention to public events of a relatively short duration, lasting just one or a few days. But these types of events are not the only potential "magnets" for sex trafficking. 
For instance, in a previous report we identified Minot, North Dakota, as a location showing large increases in sex ad volume over multiple months in 2013 using similar methods of analysis~\cite{dubrawski2015leveraging}. Around that time, the oil boom was taking place in that particular area of the state. It was apparently closely followed by a boom in the volume of sex ads published locally, and the elevated activity has persisted since.  
The presented methodology can be used to detect the onset of temporary or more persistent changes of sex advertising activity alike. 

Our work aims to shed light on the importance of data-driven approaches to understanding, characterizing, and approaching the relationship between public events (transient or persistent) and the problem of sex trafficking.  
Lack of data-driven approaches may result in a reduced awareness and suboptimal allocation of crucial resources from 
key stakeholders and result in lower than possible effectiveness of efforts to combat human trafficking crimes and rescue their victims.  
Reliance on quantitative evidence accessible through data-driven analysis can inform wise resource allocation, guide good policies, and foster the most meaningful impact.  

\Urlmuskip=0mu plus 1mu
\bibliographystyle{plain}
\bibliography{SuperBowlEventStudy}

\subsection*{Acknowledgements}
This work has been partially supported by the National Institute of Justice (2013-IJ-CX-K007),
Defense Advanced Research Projects Agency (FA8750-14-2-0244), and National Science Foundation
(1320347). The authors wish to express their gratitude to colleagues Donghan Wang and Shari Ann Cheng for their extensive help with experiments reported in this manuscript. 
\end{document}